\documentclass{elsart}

 \usepackage{graphicx}
 \usepackage{amssymb}

\begin{document}

 \begin{frontmatter}

\title{Band structure parameters of the nitrides: The origin of the small band gap of InN}
\author[label1]{Su-Huai Wei},
\author[label1]{Pierre Carrier} 
\address[label1]{National Renewable Energy Laboratory, Golden, Colorado 80401}

\begin{abstract}
Using a band-structure method that includes the
correction to the band gap error in the local density approximation
(LDA), we study the chemical trends of the band gap variation in III-V
semiconductors and predict that the band gap for InN is 0.8 $\pm$ 0.1
eV, which is much smaller than previous experimental value of $\sim
1.9$ eV.  The unusually small band gap for InN is explained in terms of
the high electronegativity of nitrogen and consequently the small
band gap deformation potential of InN. The possible origin of the 
measured large band gaps is discussed in terms of the non-parabolicity of
the bands and the Moss-Burstein shift.
Based on the error analysis of our LDA calculation and available
experimental data we have compiled the recommended band structure 
parameters for wurtzite AlN, GaN and InN.

\end{abstract}

 \end{frontmatter}

Contacting information: \\ \\
Dr. Su-Huai Wei \\
National Renewable Energy Laboratory \\
1617 Cole Blvd. \\
Golden, CO 80401, USA \\
Tel: (+1) 303-384-6666; Fax: (+1) 303-384-6432 \\
email: swei$@$nrel.gov

\newpage

\section{Introduction}

III-nitrides are usually considered as wide-band gap materials that
have applications in devices such as ultraviolet/blue/green
light-emitting diodes and lasers \cite{Morkoc}. However, recent measurements
suggest that the band gap of wurtzite (WZ) InN is below 1.0 eV \cite{Inushima,Davydov,Wu,WuBIS,Yamamoto},
much smaller than the 1.89 eV band gap \cite{Tansley} widely accepted in the past
to interpret experimental data \cite{Morkoc} and to fit empirical
pseudopotentials for modeling InN and related alloy properties \cite{Bellaiche,Kent}. If
InN indeed has a less than 1.0 eV band gap, which is even smaller than
that for InP (1.4 eV) \cite{Madelung}, then InN and its III-nitride alloys could also be
suitable for low band gap device applications such as future-generation
solar cells because the nitride alloys can cover the whole solar spectrum range. 
 \begin{figure}[h]
  \begin{center}
  \includegraphics*[width=12cm]{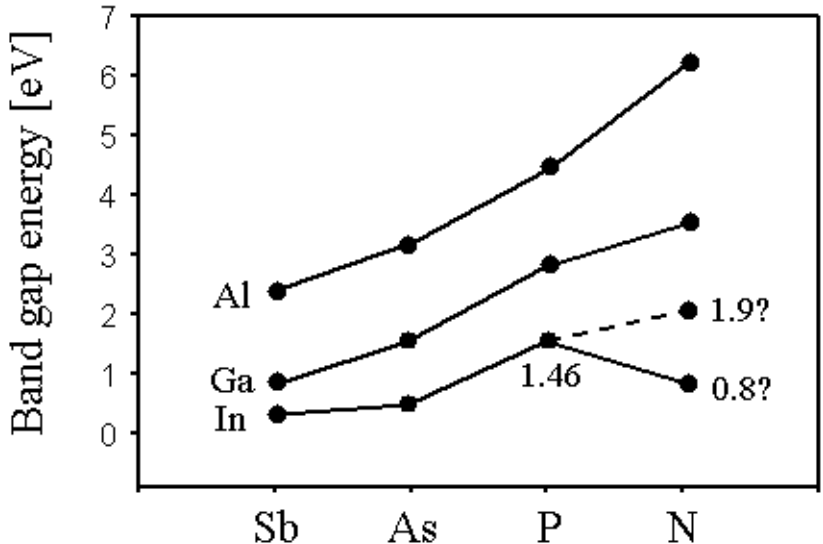}
  \end{center}
 \caption{
 Band gap as a function of anions for III-V semiconductors.
 See text for details.
  }
 \label{Fig1}
 \end{figure}

The possible low band gap of InN also provides a challenge to
understand the general chemical trends of semiconductor band gaps \cite{Nag,Wei}.
Conventional wisdom holds that for common-anion (or cation) III-V
semiconductors, the direct band gap at $\Gamma$ increases as the
cation (or anion) atomic number decreases [the band gap-common-anion
(or cation) rule].  This observation is strongly supported by
experimental data \cite{Madelung} shown in Figure 1.  For example, the direct
band gaps of the common-anion compounds InAs, GaAs, and AlAs increase
from 0.42 to 1.52 to 3.13 eV. Similarly, the direct band gaps of the
common-cation zinc-blende (ZB) compounds GaSb, GaAs, GaP, and GaN
increase from 0.81 to 1.52 to 2.86 to 3.32 eV. This trend also would
hold for common-cation InX (X=N, P, As, Sb) compounds if $E_g({\rm
InN})= 1.9$ eV, as previously reported \cite{Tansley}. However, the rule will
be broken if $E_g({\rm InN}) \sim 0.8$ eV as reported in recent
measurements \cite{Inushima,Davydov,Wu,WuBIS,Yamamoto}.

 \begin{table}[bt] 
 \caption{LDA calculated structural parameters of AlN, GaN, and InN. Results are compared with available experimental
 data (in parenthesis). $\Delta E_{ZB-WZ}$ is the calculated total
 energy difference between the ZB and WZ phases. Positive number indicate the WZ structure is more stable, in
 agreement with experiment.
 }
 \begin{center}
 \begin{tabular}{cccc}
 properties &AlN &GaN &InN \\
 \hline\hline
 $a$ (\AA)&3.098 (3.112)& 3.170 (3.189)& 3.546 (3.544)\\
 $c/a$ (\AA)&1.601 (1.601)&1.625 (1.626)&1.612 (1.613)\\
 $u$ & 0.3819 & 0.3768 & 0.3790 \\
 $a_{ZB}$ (\AA) & 4.355 (4.36)& 4.476 (4.50) & 4.964 (4.98)\\
 $\Delta E_{ZB-WZ}$ (meV/2-atom) & 45 & 11 & 21 \\
 \hline\hline
 \end{tabular}
 \end{center}
 \end{table}

Direct theoretical calculation of the band gap of InN is not straightforward, 
mainly because in modern band structure calculations employing
the local density approximation (LDA) \cite{Hohenberg}, the calculated semiconductor
band gap is severely underestimated \cite{Zhu,Rubio}.  For example, the
LDA-calculated band gap of GaAs ($\sim 0.1$ eV) is much
smaller than the experimental value of 1.52 eV. For InN
in the WZ structure, the LDA-calculated band gap is about -0.4 eV. This
value is clearly much smaller than the true band gap of InN. Various
approaches such as the GW and the self-interaction correction (SIC) methods
have been tried in the past to correct the LDA band gap error \cite{vanSchilfgaarde,Kotani,Vogel,Johnson}.  Although
there are strong indications from recent calculations \cite{Wei,Bechstedt} that the
true band gap of InN should be much smaller than the previously
reported experimental value of 1.9 eV, the uncertainty of these
calculations is still quite large. Depending on the different treatments,
the predicted band gap values for InN varies from 0.0 eV to 1.8 eV \cite{vanSchilfgaarde,Kotani,Vogel,Johnson}. 
This is partly
because the presence of the In 4$d$ orbitals in the valence bands and
partly because the LDA calculated band gap for InN is negative \cite{Bechstedt}.
 \begin{table}[bt] 
 \caption{Fitted parameters $\bar V$, $V_0$, and $r_0$ for group III and group V atoms.
 ES denotes empty sphere. For nitrides, $R_{MT}(ES)=1.68$ a.u.. For all other compounds,
 $R_{MT}(ES)=2.05$ a.u..
 }
 \begin{center}
 \begin{tabular}{cccc}
 Atom & $\bar V$ (Ry) & $V_0$ (Ry) & $r_0$ (a.u)\\
 \hline\hline
 N, P, As, Sb & 0.00 & 80 & 0.025 \\
  Al & 0.00 & 360 & 0.025 \\
  Ga & 0.00 & 280 & 0.025 \\
  In & 0.00 & 200 & 0.025 \\
  $ES$ & 0.36 & 100 & 0.025 \\
 \hline\hline
 \end{tabular}
 \label{table1}
 \end{center}
 \end{table}

\begin{table}[bt]
\caption{Calculated band gaps at $\Gamma$ for ZB
and WZ III-V compounds at experimental (exp) lattice constants using the
LDA-plus-correction (LDA+C) methods. The $E_g^{LDA+C}$ values with an (*)
are fitted values, whereas all the others are predicted values.
Our calculated results are compared with available experimental data \cite{Madelung}.
The last column show the error between predicted values and the experimental data.}
\begin{center}
\begin{tabular}{cccccc}
 & $a_{exp}$ (\AA) & $E_g^{LDA+C}$ (eV) & $E_g^{exp}$ (eV) & $|\delta E_g|$ (eV) \\
\hline\hline
AlSb & 6.133  & 2.28 &  2.32 & 0.04 \\
GaSb & 6.096  & 0.81 &  0.81 & 0.00 \\
InSb & 6.479  & 0.15 &  0.24 & 0.09 \\
AlAs & 5.660  & 3.05 &  3.13 & 0.08 \\
GaAs & 5.653  & 1.43 &  1.52 & 0.09 \\
InAs & 6.058  & 0.36 &  0.42 & 0.06 \\
AlP  & 5.467  & 4.42$^*$ & 4.38 & 0.04 \\
GaP  & 5.451  & 2.86$^*$ & 2.86 & 0.00 \\
InP  & 5.869  & 1.40$^*$ & 1.46 & 0.06 \\
AlN  & 4.360  & 6.00 &  --- &--- \\
GaN  & 4.500  & 3.34$^*$ & 3.32 & 0.02 \\
InN  & 4.980  & 0.70 &  --- & --- \\
          &a=3.112&&&&\\
AlN(WZ)   &c=4.982 & 5.95 & ~6.1 & 0.15\\
          &u=0.3819&&&&\\
          &a=3.189&&&&\\
GaN(WZ)   &c=5.185 & 3.49 & ~3.5 & 0.01\\
          &u=0.3768&&&&\\
          &a=3.544&&&&\\
InN(WZ)   &c=5.718 & 0.85 & --- & ---\\
          &u=0.3790&&&&\\
\hline\hline
\end{tabular}
\end{center}
\end{table}

In this paper, to predict the InN band gap and understand the origin of
the InN band gap anomaly, we have performed LDA-based band-structure
calculations using a semiempirical method in which the LDA band gap
error are corrected \cite{Wei}.  We find that the band gap of WZ InN is 0.8
$\pm$ 0.1 eV, in good agreement with recent experimental measurements.
We show that the reason that InN has a smaller band gap than InP is due to
the much large electronegativity and the much smaller
band gap deformation potential for InN. The semiempirical
approach is also applied to analyze the LDA error in the calculated band
structure parameters. Based on this analysis, more realistic band structure
parameters for the III-nitrides are recommended.
\begin{table}[bt]
\caption{Calculated (LDA+C) direct band gaps (in eV) at $\Gamma$ of 
zinc-blende Al, Ga, and In compounds
at their equilibrium (eq) lattice constants and at their 
respective phosphides lattice constants.} 
\begin{center}
\begin{tabular}{ccccccccc}
 & $a=a_{\rm AlP}$ & $a_{eq}$ & & $a=a_{\rm GaP}$ & $a_{eq}$  & & $a=a_{\rm InP}$ & $a_{eq}$ \\
\hline\hline
AlN & 0.45 & 6.00 & GaN & -0.61 & 3.34 & InN & -1.27 & 0.70 \\
AlP & 4.42 & 4.42 & GaP &  2.86 & 2.86 & InP &  1.40 & 1.40 \\
AlAs& 4.04 & 3.05 & GaAs&  2.36 & 1.43 & InAs&  0.92 & 0.36 \\
AlSb& 5.69 & 2.28 & GaSb&  3.67 & 0.81 & InSb&  2.15 & 0.15 \\
\hline\hline
\end{tabular}
\end{center}
\end{table}

\section{Method of calculations}

The band structure calculations in this study are performed using the
fully relativistic (including spin-orbit coupling), 
general potential, linearized augmented plane wave (LAPW)
method \cite{WeiBIS}. Highly converged k-points sampling 
for the Brillouin zone integration and cut-off energy for the basis
function are used.  The Ga 3$d$ and In 4$d$ states are treated as
valence electrons. The band structures are calculated at experimental
lattice constants \cite{Madelung}. For the III-nitride compounds, the
LDA calculated structural parameters (Table I) 
are in very good agreement with the available experimental data. 
The absorption coefficients for the nitrides 
are calculated using the optical package in WIEN2K \cite{Blaha}. 
\begin{table}[b]
\caption{Calculated atomic $s$ and $p$ orbital energies for group III and group V elements.}
\begin{center}
\begin{tabular}{ccc}
Atom & $\epsilon_s$ (eV) & $\epsilon_p$ (eV)\\
\hline\hline
Al & -7.91 & -2.86 \\
Ga&-9.25&-2.81\\
In&-8.56&-2.78\\
N&-18.49&-7.32\\
P&-14.09&-5.68\\
As&-14.70&-5.34\\
Sb&-13.16&-5.08\\
\hline\hline
\end{tabular}
\label{tab4}
\end{center}
\end{table}

 \begin{figure}[h]
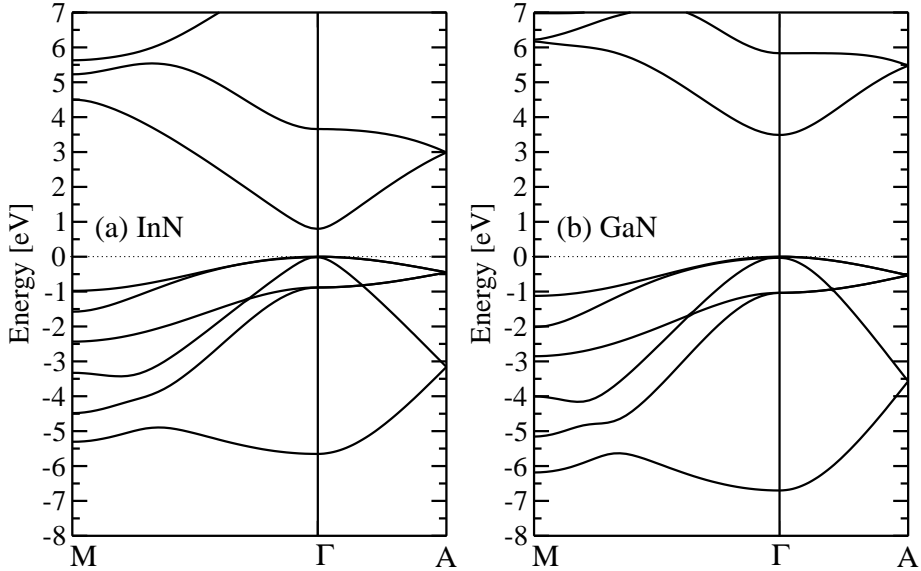

  \begin{center}
  \includegraphics*[width=6cm]{bandInN.eps}
  \includegraphics*[width=6cm]{bandGaN.eps}
  \end{center}
 \caption{
Calculated band structure of (a) wurtzite InN and (b) wurtzite GaN.
The energy zero is set at valence band maximum (VBM).
  }
 \label{Fig2}
 \end{figure}

Although LDA is accurate in predicting the ground state properties
such as the lattice parameters, it is well known that it severely
underestimates the semiconductor band gap \cite{Zhu}. To correct the LDA band gap
error, we use a well established approach by adding to the LDA
potential $\delta$-function-like external potentials \cite{Christensen,WeiTHIRD} inside the
muffin-tin (MT) spheres centered at each atomic site $\alpha$
\begin{equation}
\label{e-1}
V^\alpha_{ext}(r)=\bar V^\alpha + V_0^\alpha({r_0^\alpha \over r}) 
e^{-({r \over r_0^\alpha})^2}~,
\end{equation}
and performed the calculation self-consistently. This functional form
of the correction potential is based on the observation that the LDA
band gap error is orbital dependent. To correct the band gap error, one
needs to have a potential that is more repulsive to the $s$ orbital
than to the $p$ orbital. Because the $p$ orbital has zero charge
density at the nuclear site, whereas the $s$ orbital has finite
density at the nuclear site, a $\delta$-like potential centered at the
nuclear site can increase the band gap. The parameters in Eq. (1) are
fitted first only to the available experimental energy levels and to the
quasiparticle energies \cite{Zhu} at {\it high-symmetry} k-points for AlP,
GaP, InP, and GaN \cite{WeiTHIRD}. To improve the fit, empty spheres centered at
tetrahedral sites are also used and a constant potential term
$\bar V^\alpha=0$ is added only at the empty sphere sites.  The MT
radii for the empty sphere are 2.05 a.u except for zinc-blende (ZB) nitrides where
we used R$_{MT}=1.68$ a.u. to avoid having overlapping MT spheres.  
The fitting parameters are given in Table II. The {\it same parameters} given 
in Table II are then used to predict the band gaps of arsenides, antimonides, and 
nitrides. To find the band gap for the wurtzite structure, we add the LDA-calculated
band gap differences between the WZ and ZB compounds to
the calculated band gaps for the ZB compound. This is done to avoid using
extra fitting parameters because a smaller empty sphere has to be used
in the wurtzite structure. Comparing to our directly calculated results
for the wurtzite structure, we find that this procedure is reliable. 
The overall uncertainty of the predicted band gap associated with this fitting
procedure is estimated to be less than 0.1 eV.
 \begin{figure}[b]
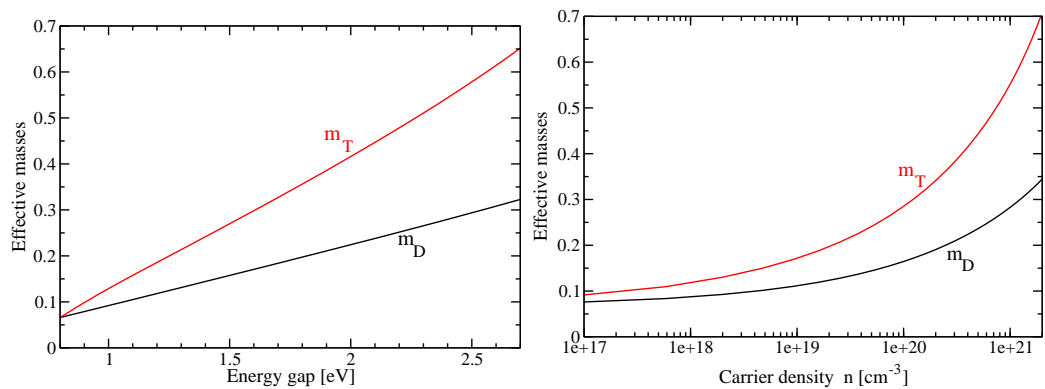

  \begin{center}
  \includegraphics*[width=6.8cm]{Eg_VS_effMass.eps}
  \includegraphics*[width=6.8cm]{Carrier_VS_effMass.eps}
  \end{center}
 \caption{
Calculated electron effective masses as a function of (a) the absorption edge energy and (b) the carrier density.
The definition of $m_D$ and $m_T$ are given in Eq. (2) and Eq. (3), respectively.
  }
 \label{Fig3}
 \end{figure}
The effective mass are calculated using two definitions. For the density
of state effective mass $m_D$ that is related to Moss-Burstein shift \cite{Moss}, 
it is given by
\begin{equation}
m_D({\bf k}) = {\hbar^2 k^2 \over {2E({\bf k})}}~.
\end{equation}
For the wurtzite compounds with anisotropic band we define
$m^*_D=[(m_D^{\perp})^2m_D^{\parallel}]^{1/3}$,
where $m_D^{\perp}$ and $m_D^{\parallel}$ are the effective masses perpendicular 
and parallel to the $c$ axis, respectively.
For transport effective mass $m_T$, it can be calculated using the definition that
\begin{equation}
m_T = {\hbar^2 k \over {dE({\bf k})/d{\bf k}}}~.
\end{equation}
The two definitions are identical at $\Gamma$ or if the band
is parabolic, but could be significantly different at large {\bf k}
when the band is non-parabolic. For most semiconductors $E({\bf k})$ contains
additional terms of the order $k^4$ \cite{Madelung}, therefore, near the conduction band minimum (CBM) the 
electron effective mass increases linearly as a function of the band edge energy $E$ and the
slop of $m_T$ versus E is about twice as larger as the slope of $m_D$ versus E.
It is well know that LDA also
underestimate the calculated effective masses. Similar to the treatment 
for the band gap, we have fitted our results for GaN and GaAs, and applied the 
same procedures to calculate the effective masses for AlN and InN.

\section{Results and discussions}

\subsection{Band gap of InN}

The predicted direct band gaps at the $\Gamma$-point for the III-V
semiconductors are shown in Table III. These values are compared with
available experimental data \cite{Madelung}.  We find that for nearly {\it all} the
III-V semiconductors, the differences between the predicted
and the experimental band gaps are less than 0.1 eV. For InN, however,
our predicted value of 0.85 eV is much smaller than the previous
experimental value \cite{Tansley} of 1.9 eV, but it is in very good
agreement with recent experimental measurements \cite{Inushima,Davydov,Wu,WuBIS,Yamamoto}. For AlN, our
predicted band gap of 6.0 eV is also close to recent photoluminescence measurement of
6.1 eV \cite{Li}, which is smaller than previously accepted value around 6.3 eV 
\cite{Morkoc,Madelung,Edgar}.

\subsection{Chemical trends of the band gap of III-V semiconductors}

Our calculations above show convincingly that the band gap of InN is
around 0.8 $\pm$ 0.1 eV. However, this value is about 0.6 eV smaller
than that of InP, thus contradicting the conventional wisdom that the
band gaps of common-cation (or anion) compounds increase as the anion
(cation) atomic number decreases (Fig. 1). To understand the origin of
the breakdown of the band gap common-cation rule in In compounds, we
study the chemical and size contributions to the band gap in III-V
semiconductors. For the chemical contribution, we calculate the
band gaps of Al, Ga, and In compounds at the {\it fixed lattice constants} of
AlP, GaP, and InP, respectively. The results are shown in Table
IV. LDA corrections (LDA+C) are included. We find that at the phosphide
volume, the band gaps of the common-cation system decrease from $M$Sb
to $M$P to $M$As to $M$N ($M$=Al, Ga, and In), following the same
trend of the anion atomic valence $s$ orbital energies shown in Table
V. This is because the conduction band minimum (CBM) at the
$\Gamma$-point is an anion $s$ plus cation $s$ state. The anion
contribution increases as the compound becomes more ionic. 
Because the N 2$s$ orbital energy is much lower in energy
than the Sb 5$s$, As 4$s$ and P 3$s$ orbital energies (Table V), respectively,
the band gap of the nitrides are also much lower than the corresponding
antimonides, arsenides, and phosphides at fixed volume.
\begin{table}[bt]
\caption{Calculated band gap volume deformation potential $a_g=dE_g/dlnV$ at $\Gamma$ for III-V semiconductors.} 
\begin{center}
\begin{tabular}{cccccc}
Compound  & $-a_{g}$ & Compound  & $-a_{g}$ & Compound  & $-a_{g}$  \\
\hline\hline
~~AlN (ZB; WZ)~~~& 10.2; 10.4 & ~~~GaN (ZB; WZ)~~~ & 7.4; 7.8 & ~~~InN (ZB; WZ)~~~ & 3.7; 4.2 \\
AlP & 9.5 & GaP &  8.8 &  InP &  5.9 \\
AlAs& 8.9 & GaAs &  8.2 & InAs &  5.7 \\
AlSb& 8.9 & GaSb &  8.0 & InSb &  6.4 \\
\hline\hline
\end{tabular}
\end{center}
\end{table}

Because the order of the band gaps calculated at the fixed volume is
generally opposite to what is observed at the equilibrium lattice
constants, the chemical contribution alone cannot explain the
experimentally observed trend in the band gaps at equilibrium lattice
constants. Next, we investigate the size or volume deformation
contribution to the band gap. The calculated volume deformation
potentials \cite{WeiFOURTH} $a_g=dE_g/dlnV$ with the LDA correction for III-V semiconductors are
listed in Table VI. We see that all the compounds have negative
volume deformation potentials at $\Gamma$, i.e., when the volume decreases,
the band gap increases. Therefore, it is clear that the observed common-cation
rule and the common-anion rule for the band gap is mainly due to the
large deformation potential of the III-V compounds. For example, at
GaP lattice constant, the band gap of GaSb is 0.81 eV larger than that
of GaP. However, GaSb is about 34\% larger in volume than GaP. So, with
an average deformation potential of -8.4 eV, the band gap of GaSb at
its equilibrium lattice constant is about 2.05 eV smaller than the
band gap of GaP at its equilibrium lattice constant. The same situation
also applies to AlN and GaN versus AlP and GaP, respectively, because AlN and GaN
have large band gap deformation potentials [$a_g({\rm AlN})=-10.4$ eV 
and $a_g({\rm GaN})=-7.8$ eV]. However, for InN, although its volume is about 49\%
smaller than InP, its band gap deformation potential is small,
$a_g({\rm InN})=-4.2$ eV. Because of this small $|a_g|$, the contribution due to
the size or deformation potential ($\sim$ 2.1 eV) is not sufficient to reverse the
order of the band gap due to the contribution of the chemical effect ($\sim$ $-$2.7 eV).
This explains why InN has a band gap about 0.6 eV smaller than that of InP.
\begin{table}[bt]
\caption{Recommended band structure parameters at $\Gamma$ for unstrained AlN, GaN, and InN. The properties show in this table are
the band gap $E_g$, the spin-orbit splitting $\Delta_0$, the crystal field splitting $\Delta_{CF}$, the valence band splittings
$\Delta E_{12}$ and $\Delta E_{13}$, and the effective masses $m$ parallel and perpendicular to the $c$ axis. The averaged
effective mass can be obtained using $m^*=[(m^{\perp})^2m^{\parallel}]^{1/3}$.
}
\begin{center}
\begin{tabular}{ccccccc}
~~~~~Properties~~~~~~ &\multicolumn{2}{c} {AlN} &\multicolumn{2}{c} {GaN} &\multicolumn{2}{c} {InN} \\
\hline\hline
$E_g$(WZ) (eV) & \multicolumn{2}{c}{6.10}&\multicolumn{2}{c}{3.51}&\multicolumn{2}{c}{0.78} \\
$E_g$(ZB) (eV) & \multicolumn{2}{c}{6.15}&\multicolumn{2}{c}{3.35}&\multicolumn{2}{c}{0.70} \\
$\Delta_0$ (meV) & \multicolumn{2}{c} {19} & \multicolumn{2}{c} {16} & \multicolumn{2}{c} {5} \\
$\Delta_{CF}$ (meV) & \multicolumn{2}{c} {$-$224} & \multicolumn{2}{c} {25} & \multicolumn{2}{c} {19} \\
~~~$\Delta E_{12}$ (meV)~~~~~ & \multicolumn{2}{c}{218} & \multicolumn{2}{c}{8} & \multicolumn{2}{c}{3} \\
$\Delta E_{13}$ (meV) & \multicolumn{2}{c}{237} & \multicolumn{2}{c}{33} & \multicolumn{2}{c}{21} \\
\hline\hline
& \multicolumn{6}{c}{effective masses} \\
  & ~~~~~~$\perp$~~~~~~ & ~~~~~~$\parallel$~~~~~~ & ~~~~~~$\perp$~~~~~~ & ~~~~~~$\parallel$~~~~~~ & ~~~~~~$\perp$~~~~~~ 
& ~~~~~~$\parallel$~~~~~~ \\
\hline\hline
$m_A (m_0)$& 4.35 & 0.28 & 0.39 & 2.04 & 0.14 & 2.09 \\
$m_B (m_0)$& 0.67 & 3.50 & 0.43 & 0.85 & 0.13 & 0.50 \\
$m_C (m_0)$& 0.68 & 3.43 & 1.05 & 0.19 & 0.81 & 0.07 \\
$m_e (m_0)$& 0.33 & 0.32 & 0.22 & 0.20 & 0.07 & 0.06 \\
\hline\hline
\end{tabular}
\end{center}
\end{table}

From the analysis above, we see that the breakdown of the
common-cation rule for the band gap in In compounds is due to the
small $|a_g|$. We find that \cite{WeiFOURTH}, the small
$|a_g|$ for InN is due to the combined effects of (i) a large difference
between the cation In $5s$ and anion N $2s$ orbital energies, (ii) a
large repulsion between the N $2p$ and the high-lying In $4d$
orbitals, and (iii) a large In-N bond length (relative to AlN and GaN). 
Because a similar situation also exists in II-VI semiconductors, 
one would expect that the breakdown of the common-cation rule should also
apply to the II-VI systems. Indeed, experimental data \cite{Madelung} show that the ZnO
band gap of 3.4 eV is smaller than the ZnS band gap of 3.8 eV. Our 
calculations also show that CdO and HgO would have band gaps that are about 0.5 eV 
smaller than the band gaps of CdS and HgS, respectively, if they could all
exist in the ZB phase. 
 \begin{figure}[h]
  \begin{center}
  \includegraphics*[width=12cm]{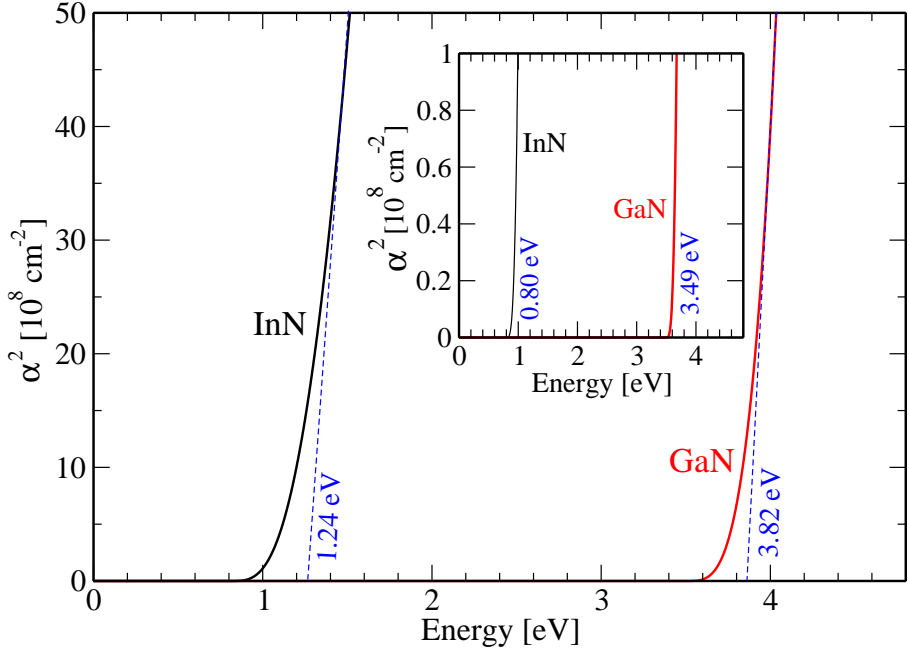}
  \end{center}
 \caption{
Calculated absorption coefficients of InN and GaN. We show that using the 
linear extrapolation technique the apparent measured band gaps depend on the scale
used in the extrapolation.
  }
 \label{Fig4}
 \end{figure}

\subsection{Possible origin of the measured large band gap for InN}

Our calculation and analysis above show that the fundamental band gap of InN
is indeed small, around 0.8 eV. To understand the origin of some of the experiments
which show large band gap of InN \cite{Tansley,Motlan,Osamura}, we have performed detailed study of 
the band structure of InN. 
Experimentally, one often assumes that the band edge states near $\Gamma$ is 
parabolic and the dipole transition matrix element is nearly independent of {\bf k}, 
therefore, the absorption coefficient squared $\alpha^2$ is a linear function
of the absorption energy $E$. The fundamental band gap, thus, can be obtained from the interception
with the energy axis by drawing a straight line in the $\alpha^2$ versus $E$ plot \cite{Tansley,Motlan,Osamura}.  
To test the validity of this assumption, we show in Figure 2 the calculated band structure of 
wurtzite InN. Figure 3 shows the calculated electron effective masses, and
Figure 4 shows our calculated absorption coefficients of InN. For comparison,
we also calculate the band structure and absorption coefficients of GaN.
We find that the conduction band of InN is strongly non-parabolic (Fig.2). This is confirmed
from the calculated electron effective masses (Fig. 3), which increase significantly with the
band edge energy or electron concentration. If the conduction band was parabolic the
electron effective mass would be a constant. Because of the large deviation from the
parabolic band, the calculated $\alpha^2$ is not a linear function of $E$ (Fig. 4). Therefore,
if one use the linear extrapolation technique to determine the band gap, the
derived apparent band gaps depend on where the straight lines are drawn. For example, as
shown in Fig. 4 for InN, using large values of the absorption coefficient to draw the straight line,
one can obtain an apparent band gap that is about 0.4 eV larger than the fundamental band gap. 
The dependence is relatively smaller for GaN which has a larger effective mass than that of InN, thus a 
larger density of states near the  CBM and a sharper increase of the absorption coefficient. 
We notice that the samples used to obtain the large InN band gaps \cite{Tansley,Motlan,Osamura} often
have poor sample quality and the band gaps are usually estimated from the absorption spectra
with large absorption coefficients. We also notice that the InN samples that show
large band gaps often have high oxygen concentration \cite{Motlan} and are heavily n-type doped 
\cite{Tansley,Motlan,Osamura}. Besides the possible formation of InNO alloys as proposed in Ref. \cite{Motlan},
this observation suggests that the measured absorption edge can be shifted by the 
Moss-Burstein effect \cite{Moss}. Figure 5 shows our calculated absorption edge energy
as a function of the carrier density. We see that the absorption edge increases with
the carrier density from 0.8 eV for intrinsic InN to $\sim$ 2.5 eV for sample with electron
concentration of $\sim$10$^{21}$ cm$^{3}$. These results are consistent with recent experimental
observation \cite{WuTHIRD}.
 \begin{figure}[h]
  \begin{center}
  \includegraphics*[width=12cm]{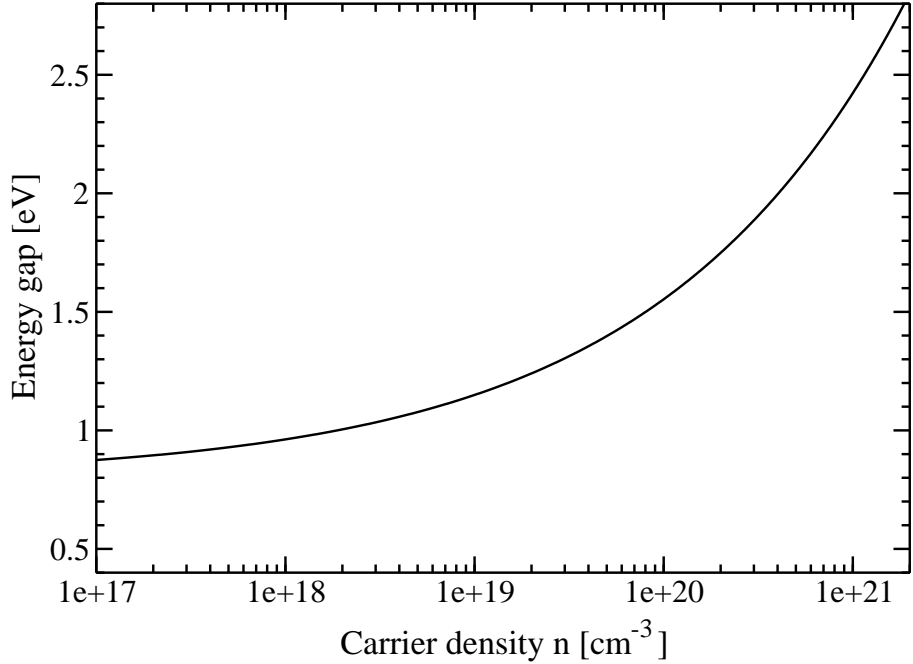}
  \end{center}
 \caption{
Calculated Moss-Burstein shift of the absorption edge energy as a function of the carrier density.
  }
 \label{Fig5}
 \end{figure}

\section{Band structure parameters of nitrides}

In the last few years several excellent review articles \cite{Vurgaftman,Ghuiyan} and books 
\cite{Morkoc,Edgar} have
been published to described the band structure parameters of InN and other III-nitride.
The recommended band structure parameters are based on a collections of available 
experimental data and theoretical calculations. The approach used in this study enabled
us to analyze the degree of the systematic LDA errors on the band structure parameters
of the III-nitrides. For example, we find that the LDA not only underestimate the band gap,
but it also slightly underestimated the crystal field splittings $\Delta_{CF}$ and the spin-orbit
splittings $\Delta_{0}$ at the top of the valence band, as well as the electron and light hole 
effective masses. Details of the analysis will be published elsewhere. Based on the LDA error analysis and 
comparison with available experimental data we propose here the recommended band structure
parameters for the III-nitrides shown in Table VII. We suggest that these parameters should 
be used in the future to fit empirical pseudopotentials to study the nitride systems.

\section{Summary}

In summary, using an empirical LDA-based band structure method with band gap correction
we have systematically studied the chemical trends of the band gap variation in III-V
semiconductors. We find that InN has a band gap of $0.8 \pm 0.1$ eV,
in good agreement with recent experimental measurements. We show that
the previously accepted band gap-common-cation rule 
does not hold for ionic InN and the II-VI oxides. We have also compiled the 
recommended band structure parameters for AlN, GaN, and InN.

\section{Acknowledgment}

We would like to thank I. G. Batyrev, X. Nie, S. B. Zhang, S. R. Kurtz, H. X. Jiang, Y. F. Chen, 
and G. X. Gao for many valuable discussions. This work is supported by the U.S. Department of Energy, 
Contract No. DE-AC36-99GO10337.


\end{document}